\documentclass{article}
\usepackage[T1]{fontenc}
\usepackage{mathrsfs}
\usepackage{amsmath}
\usepackage{amssymb}
\usepackage[dvipdfm]{graphicx}
\usepackage{tabularx}
\usepackage{bm}

\def\abstract#1{\vskip 7mm 
        \begin{center}{\large Abstract}\par \smallskip
                \begin{minipage}[c]{12cm}
                        \small #1
                \end{minipage}
        \end{center}
}
\def\title#1{\begin{center}{\Large\bf #1}\end{center}}
\def\author#1{\vskip 5mm \begin{center}{#1}\end{center}}
\def\address#1{\begin{center}{\it #1}\end{center}}
\makeatletter
\@ifundefined{lesssim}{\def\lesssim{\mathrel{\mathpalette\vereq<}}}{}
\@ifundefined{gtrsim}{}{}
\def\vereq#1#2{\lower3pt\vbox{\baselineskip1.5pt \lineskip1.5pt
\ialign{$\m@th#1\hfill##\hfil$\crcr#2\crcr\sim\crcr}}}
\makeatother

\begin{document}

\title{%
  Evaporating (2+1)-dimensional black strings
}
\author{%
  Keiju Murata\footnote{E-mail:murata@tap.scphys.kyoto-u.ac.jp}, 
  Jiro Soda\footnote{E-mail:jiro@tap.scphys.kyoto-u.ac.jp}
  and
  Sugumi Kanno\footnote{E-mail:sugumi@hep.physics.mcgill.ca}
}
\address{%
  $^{1,2}$Department of Physics,  Kyoto University, Kyoto 606-8501, Japan\\
  $^3$Department of Physics,  McGill University, Montr\'{e}al, QC H3A 2T8, Canada
}

\abstract{
  We investigate (2+1)-dimensional black strings
  in the Kaluza-Klein spacetime. The system is classically stable
  as long as the horizon size is much larger than the size of the
  compact space. Semiclassically, however, the horizon size shrinks
  gradually due to the energy loss through the Hawking radiation.
  Eventually, the system will enter into the regime of the
  Gregory-Laflamme instability and get destabilized. Subsequently,
  the spherically symmetric black hole is formed and evaporated in the usual manner.
  This standard picture
  may be altered by the dynamics of the internal space which induced by the Hawking
  radiation. We argue that the black string is excised from the Kaluza-Klein
  spacetime before the onset of the Gregory-Laflamme instability and therefore before the 
  evaporation.
}

\section{Introduction}

In the conventional 4-dimensional spacetime,
the black hole is considered as the blackbody which emits the Hawking radiation. 
Because of this radiation, the size of the horizon gradually shrinks
and finally the black hole evaporates. Although there still exists debate
on the final state after evaporation, the total picture is simple.

\begin{figure}[h]
\begin{center}
\includegraphics[height=6cm,clip]{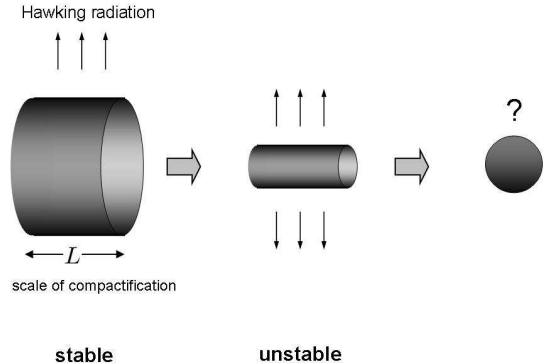}
\caption{\label{fig:naiveKKBHevap} A naive picture of the evaporation process of the
 black string. Because of the Gregory-Laflamme instability, the black string becomes the spherically 
 symmetric black hole. }
\end{center}
\end{figure}

From the point of view of the super string theory,  it is natural to consider
higher dimensional spacetime. In this picture, the event horizon of the black hole would be
the direct product of the usual event horizon  and the compact internal space.
It is interesting to investigate the evaporation process of the black hole 
in the Kaluza-Klein spacetime. In this paper, for simplicity,
we consider  the (2+1)-dimensional spacetime with $S^1$ as the compact internal space.
The resulting Kaluza-Klein black hole is called the black string.
The evaporation processes of the black strings could be different from
the 4-dimensional one due to the Gregory-Laflamme instability~\cite{Gregory:1993vy}. 
It is known that the black string is unstable when the horizon radius is
smaller than the scale of compactification. This
instability changes the spacetime structure. The evaporation
process taking into account this instability is
depicted in the Figure \ref{fig:naiveKKBHevap}.
Soon after the onset of the Gregory-Laflamme instability, the spherically symmetric
 black hole would be formed.
Subsequent evaporation process is very similar to the standard one.

Previously, we have considered the interplay between the radion and 
the Gregory-Laflamme instability~\cite{Kanno:2003au,Tamaki:2003bq,Kanno:2005mv}.
There, the radion has played an important role.
It is natural to expect the radion also makes a significant contribution
to the evaporation process of the black strings.
 To the best of our knowledge,
however, the interplay between the Hawking radiation and 
the radion dynamics has not been considered at all.
Here, we study the role of the radion dynamics
in the evaporation process of the (2+1)-dimensional black string.  

The back reaction of the Hawking radiation could destabilize the radion.
  In fact, the energy of the Hawking radiation attracts the space around it.
 Thus, the radion is deformed by the Hawking radiation. 
 In general, this deformation would be inhomogeneous. Hence, the spacetime is pinched
 at some radius. Consequently, assuming the singularity resolution, 
 the black string might be excised from the spacetime. 
If this speculation is true, the radion dynamics would change the naive
evaporation process completely.
It should be stressed that the black string excision from the spacetime
could occur  before the onset of the Gregory-Laflamme instability and hence
before the evaporation through the conventional process.

\section{(2+1)-dimensional Black Strings}

In this section, we present our model and show the black string
exists in this simple set up. 
We start with the (2+1)-dimensional dilaton gravity of the form
\begin{equation}
 S[A,g] = M_3 \int d^3x \sqrt{-G} [A R^{(3)} + \frac{\lambda^2}{A}]\ ,
\label{eq:dilaton_a=0}
\end{equation}
where $G_{\mu\nu}$ is the 3-dimensional metric, $R^{(3)}$ is the 3-dimensional
scalar curvature, 
$A$ is the dilaton field, $M_3$ is the
3-dimensional Planck mass and $\lambda$ is the parameter which have the
mass dimension. The solution of the equation of motion obtained from the action
(\ref{eq:dilaton_a=0}) is
\begin{eqnarray}
 A = \lambda r\ ,\quad
 ds^2 = -\alpha(r)dt^2 + \alpha(r)^{-1}dr^2 + dy^2\ ,\quad
 \alpha(r) \equiv \ln\left(\frac{r}{r_H}\right) \ ,
\end{eqnarray}
where the period of the internal coordinate $y$ is taken to be $L$.
This solution represents the black string with horizon radius $r_H$ and this
spacetime is asymptotically flat because the curvature reads $R = 1/r^2$.
 It is known that the black string is unstable when $r_H \lesssim L$. 
 The other way around, the black hole is stable as long as $r_H$ is much larger than $L$.
 It is common to ignore the radion dynamics when we  discuss this classical instability.
 However, once we take into account the semiclassical effect, we cannot
 ignore the radion dynamics. 
 We need to study the evaporation process of this black string
 with the non-trivial radion dynamics. 

\section{Back Reaction of the Hawking Radiation}

We consider  quantum effect of the scalar field $f$ in the black string background
with the action
\begin{equation}
 S[A,g,f] = M_3 \int d^3x \sqrt{-G} [AR^{(3)} + \frac{\lambda^2}{A}] + \int
  d^3x \sqrt{-G}[-\frac{1}{2}(\nabla f)^2]\ .
\label{eq:dilaton_a=0+matt}
\end{equation}
To take into account the radion dynamics,  we parametrize the metric as
\begin{equation}
 ds^2 = g_{ab}(x^a)dx^a dx^b + e^{-2\chi(x^a)}dy^2\ ,
\end{equation}
where $\chi$ is the radion field.
Here, we assume $r_H \gg L$ and all fields do not depend on $y$. 
Then, we can carry out the $y$ integration in
the action (\ref{eq:dilaton_a=0+matt}) to obtain
\begin{equation}
 S[A,g,\chi,f] = M_3 L \int d^2x \sqrt{-g}\,e^{-\chi}[AR-2\nabla A \cdot
  \nabla \chi + \frac{\lambda^2}{A}] + \int
  d^2x \sqrt{-g}\,e^{-\chi}[-\frac{1}{2}(\nabla f)^2]\ ,
\end{equation}
where $R$ represents the 2-dimensional scalar curvature.
We can study the evaporation process of the black string using the above
effective 2-dimensional dilaton gravity. We treat the scalar field quantum
mechanically, because we want to study the back reaction of the Hawking
radiation. We define  the effective action
\begin{equation}
 W[A,g,\chi] = -i\ln\left(\int \mathscr{D}f\,\exp(i \int
  d^2x \sqrt{-g}\,e^{-\chi}[-\frac{1}{2}(\nabla f)^2])\right)
\end{equation}
and semiclassical energy momentum tensor
\begin{equation}
 \langle T_{ab} \rangle = \frac{-2}{\sqrt{-g}}\frac{\delta W}{\delta g^{ab}}\ .
\end{equation} 
We can treat the back reaction perturbatively, that is, we regard
 the source $T_{ab}$ as the small quantity.
 The master equation for the perturbed radion $\delta\chi$ is given by
\begin{equation}
 -\Box \delta\chi - \frac{1}{A}\nabla A \cdot \nabla\delta\chi =
  \frac{1}{4M_3 L} \frac{1}{A}\langle T^a_a \rangle\ .
\label{eq:master}
\end{equation}
Note that the perturbed radion is gauge invariant as $\chi=0$. Therefore, the gauge mode cannot
appear in the master equation (\ref{eq:master}). Since the classical action of the
matter field is Weyl invariant, the trace part of the energy-momentum tensor 
$T^a_a$ should be zero. As is well known, however, Weyl symmetry has the anomaly 
in the quantum theory, namely,
\begin{equation}
 \langle T^a_a \rangle = \frac{1}{24\pi}R = \frac{1}{24\pi r^2}.
\end{equation}
Now, we can discuss the radion dynamics. 
\begin{figure}[h]
\begin{center}
\includegraphics[height=6cm,clip]{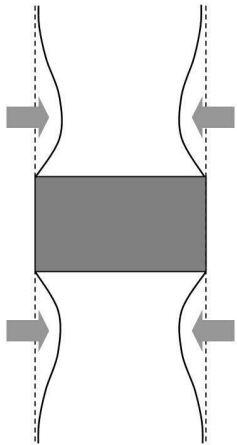}
\caption{\label{fig:radion_dynamics_onset}The radion has inhomogeneous
profile due to the Hawking flux.}
\end{center}
\end{figure}

\section{Excision of Black Strings}

We set the initial conditions 
$\delta\chi(t=0,r)=\delta\chi_{,t}(t=0,r)=0$. Then, we can deduce the radion dynamics
for small $t$ from master equation (\ref{eq:master});
\begin{equation}
 \delta\chi = \frac{1}{8M_3 L} \frac{\alpha(r)}{A}\langle T^a_a \rangle
  \, t^2 = \frac{1}{192\pi\lambda M_3 L} \frac{\ln(r/r_H)}{r^3}\, t^2\ .
\label{eq:radion_dynamics}
\end{equation}
 We gave a schematic picture of the radion dynamics in the Figure
 \ref{fig:radion_dynamics_onset}.

\begin{figure}[h]
\begin{center}
\includegraphics[height=6cm,clip]{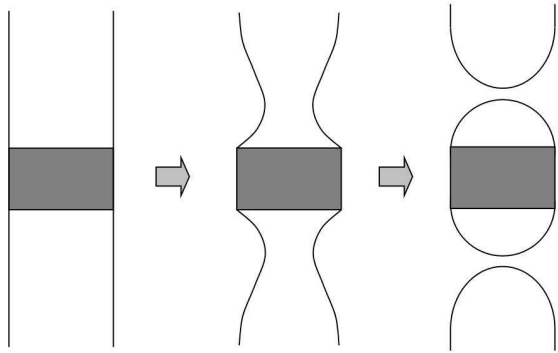}
\caption{\label{fig:radion_dynamics_nonlinear}The speculative
 non-linear radion dynamics is shown. The black string is excised from the spacetime
 at the end of the day. }
\end{center}
\end{figure}

Now, we shall speculate the non-linear radion dynamics. The radion
dynamics Eq.(\ref{eq:radion_dynamics}) can be understood from analogy to the cosmic
dynamics. In cosmology, the matter tends to shrink the space
and, finally, makes the singularity. Similarly, the internal
space may collapse due to the Hawking flux. Moreover, 
this occurs inhomogeneously. So we expect that the internal space  is pinched
off due to the Hawking radiation (Figure \ref{fig:radion_dynamics_nonlinear}).
Though the pinched point is classically singular, 
this is expected to be regularized by quantum effect.
Thus, the black string region is excised from our spacetime.
It looks like the evaporation of the black string from the observer outside.
 However, this evaporation process is completely different from 
 the naive evaporation process in Figure \ref{fig:naiveKKBHevap}.

\section{Conclusion}

 We have investigated the (2+1)-dimensional black strings
  in the Kaluza-Klein spacetime. 
 The solution of the master
equation tells us that the internal space shrinks inhomogeneously (Figure
\ref{fig:radion_dynamics_onset}). From this result, we can give the
speculation for the non-linear radion dynamics. 
The black string may be excised from the spacetime
 due to the Hawking radiation (Figure \ref{fig:radion_dynamics_nonlinear}). 
 This can be interpreted as the evaporation of the black string from the observer outside.
 This evaporation process is different from that naively considered
so far. Of course, at the present level of the analysis, we can not
insist it strongly. 
However, it is worth to reconsider the evaporation process of black strings
with taking into account the radion dynamics~\cite{Murata}.

We need to analyze the non-linear radion dynamics to examine if our expectation
 in Figure \ref{fig:radion_dynamics_nonlinear} is correct. 
Furthermore, it is intriguing to consider the present issue in the context of the
string theory.

\vskip 0.2cm
J.S. is supported by  
the Japan-U.K. Research Cooperative Program, the Japan-France Research
Cooperative Program,  Grant-in-Aid for  Scientific
Research Fund of the Ministry of Education, Science and Culture of Japan 
 No.18540262 and No.17340075.  
S.K. is supported by JSPS Postdoctoral Fellowships
for Research Abroad.

\end{document}